\documentclass[Physsubmission, Phys,a4paper]{SciPost}

\usepackage{hyperref}
\hypersetup{
    colorlinks,
    linkcolor={red!50!black},
    citecolor={blue!50!black},
    urlcolor={blue!80!black}
}

\usepackage[bitstream-charter]{mathdesign}
\DeclareSymbolFont{usualmathcal}{OMS}{cmsy}{m}{n}
\DeclareSymbolFontAlphabet{\mathcal}{usualmathcal}

\newcommand{\sigmadip}{{ \sigma_{\textnormal{dip}} }}

\newcommand{\ud}{\, \textnormal{d}}
\newcommand{\rt}{{\mathbf{r}_T}}
\newcommand{\bt}{{\mathbf{b}_T}}
\newcommand{\xt}{{\mathbf{x}_T}}

\newcommand{\ampli}{{\mathcal{N}}}
\newcommand{\Deltat}{{\boldsymbol{\Delta}_T}}
\newcommand{\nc}{{N_\textnormal{c}}}
\newcommand{\as}{\alpha_{\textnormal{s}}}

\newcommand{\pt}{{\mathbf{p}_T}}

\newcommand{\ktt}{{k_T}}

\usepackage{etoolbox}
\apptocmd{\thebibliography}{\setlength{\itemsep}{0pt}}{}{}

\usepackage{tikz}
\usepackage{wrapfig}

\begin{document}

\begin{center}{\Large \textbf{
Ultraperipheral collisions and low-$x$ physics
}}\end{center}

\begin{center}
T. Lappi\textsuperscript{1,2$\star$}
\end{center}

\begin{center}
{\bf 1} Department of Physics, University of Jyv\"askyl\"a %
 P.O. Box 35, 40014 University of Jyv\"askyl\"a, Finland
\\
{\bf 2} 
Helsinki Institute of Physics, P.O. Box 64, 00014 University of Helsinki, Finland
\\
* tuomas.v.v.lappi@jyu.fi
\end{center}

\begin{center}
\today
\end{center}


\definecolor{palegray}{gray}{0.95}
\begin{center}
\colorbox{palegray}{
  \begin{tabular}{rr}
  \begin{minipage}{0.1\textwidth}
    \includegraphics[width=22mm]{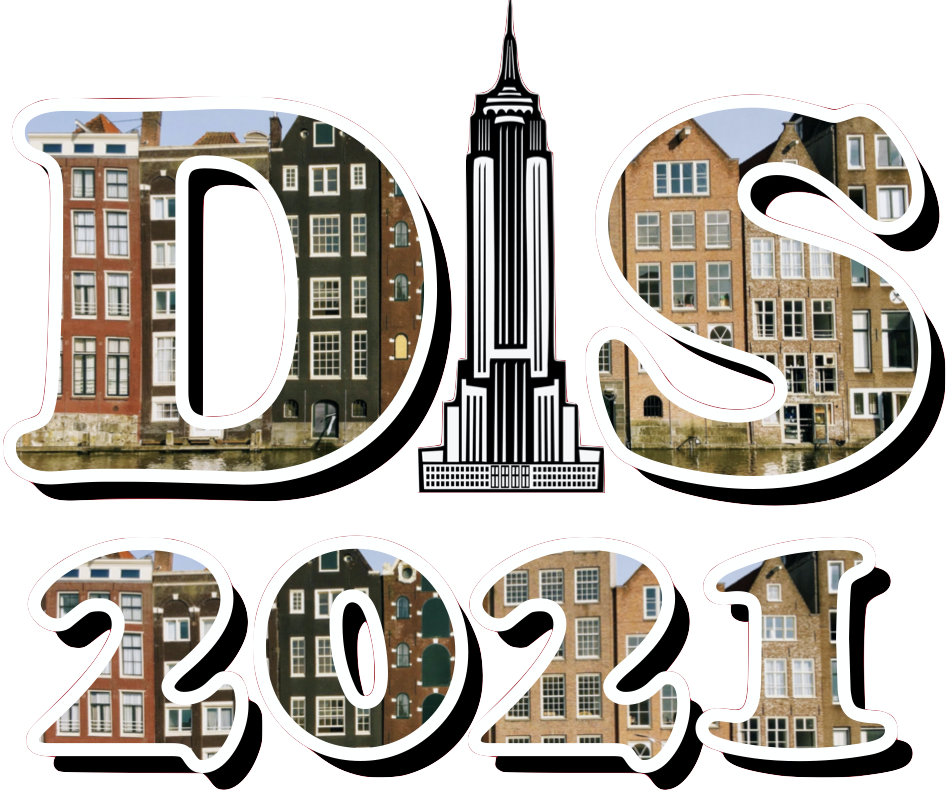}
  \end{minipage}
  &
  \begin{minipage}{0.75\textwidth}
    \begin{center}
    {\it Proceedings for the XXVIII International Workshop\\ on Deep-Inelastic Scattering and
Related Subjects,}\\
    {\it Stony Brook University, New York, USA, 12-16 April 2021} \\
    \doi{10.21468/SciPostPhysProc.?}\\
    \end{center}
  \end{minipage}
\end{tabular}
}
\end{center}

\section*{Abstract}
{\bf 
Ultraperipheral collisions at the LHC and RHIC offer the highest currently available energy for photon-nucleon and photon-nucleus collisions. Thus they are a valuable tool for studying the gluonic structure of hadrons and nuclei at small $x$.  We  discuss recent theoretical work towards understanding such exclusive processes at NLO accuracy in QCD perturbation theory. These theoretical advances are also immediately relevant for understanding the physics of deep inelastic scattering at small $x$. We also discuss experimental results in ultraperipheral collisions, most prominently for exclusive vector meson production.
}

\vspace{10pt}
\noindent\rule{\textwidth}{1pt}
\tableofcontents\thispagestyle{fancy}
\noindent\rule{\textwidth}{1pt}
\vspace{10pt}

\section{Introduction}

In the expectation of a new era of high energy deep inelastic scattering (DIS) experiments with the EIC~\cite{AbdulKhalek:2021gbh}, ultraperipheral collisions of heavy ions~\cite{Baltz:2007kq} offer a way to study some of the same physical processes at existing colliders. The electromagnetic field of a fully ionized heavy nucleus traveling at close to the speed of light can, using the Weizsäcker-Williams equivalent photon approximation, be understood as a flux of high energy quasi-real ($Q^2\approx 0$) photons. The photon flux and polarization structure can be calculated to a rather good accuracy. This makes it possible to study high energy photon-nucleus, photon-proton and photon-photon collisions at the LHC and RHIC.

Here, we will focus on the small-$x$ physics and saturation aspects of these collisions. We will first describe the dipole picture of DIS and the Color Glass Condensate (CGC) effective theory for QCD in the saturation limit. We will then discuss the complemetary description of exclusive processes using the collinear factorization approach. We then discuss some recent experimental measurements in ultraperipheral collisions, both for exclusive vector meson production and for other UPC measurements.

\section{Dipole picture of DIS and the Color Glass Condensate}

\begin{wrapfigure}{R}{0.4\textwidth}
\centerline{
\includegraphics[width=5.4cm]{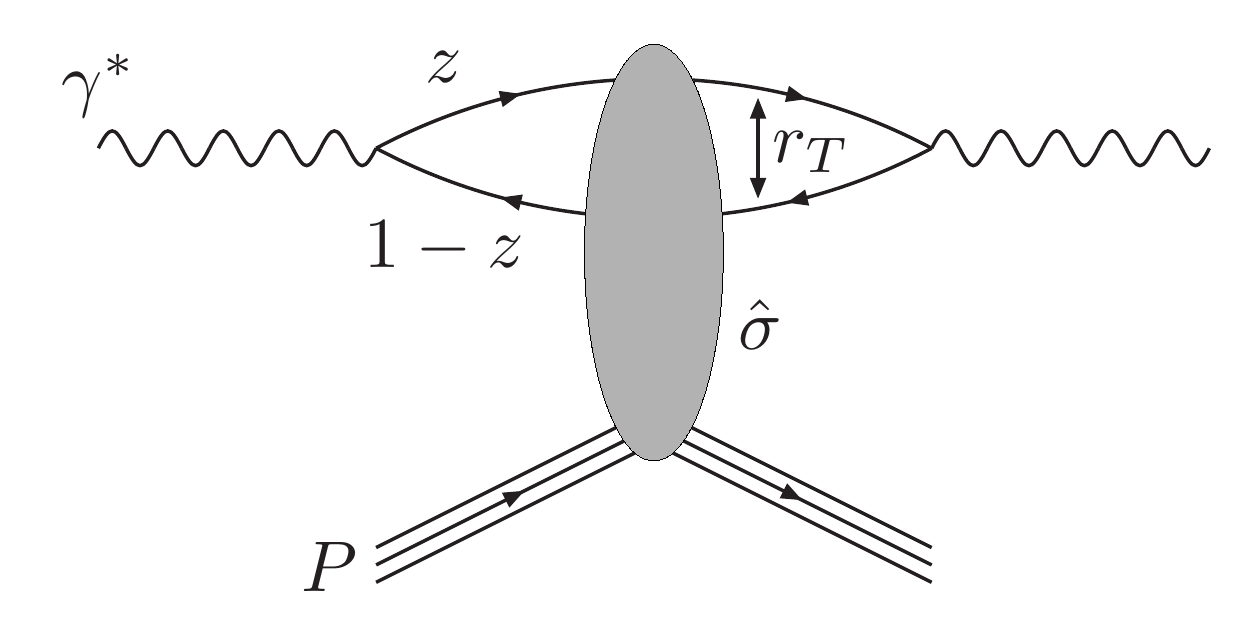}
\begin{tikzpicture}[overlay]
\node at (-2.55cm,0.2cm) {$t=-\Deltat^2$};
\draw[line width=2pt] (-2.3cm,0.65cm) arc (-30:-150:0.3cm);
\end{tikzpicture}
}
\caption{The dipole picture of DIS.}
\label{fig:dipole}
\end{wrapfigure}
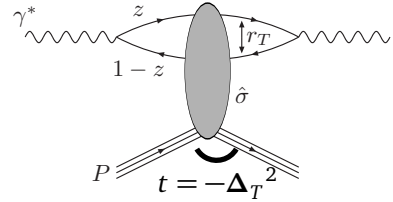
To understand the dipole picture of DIS, one should place oneself in the rest frame of the target (or more accurately in the ``dipole frame''~\cite{Mueller:2001fv} where the photon energy is larger than the virtuality, but the target has most of the collision energy). In such a frame the photon develops a cloud of QCD Fock state components which, at high collision energy, then interact instantaneously with the gluonic shockwave of the target. The leading such QCD Fock state is a color neutral quark-antiquark dipole, hence the term ``dipole picture'', see Fig.~ \ref{fig:dipole}.

The information on the small-$x$, predominantly gluonic, degrees of freedom of the target is contained in the so called ``dipole cross section.'' More properly speaking the degree of freedom is the  elastic forward scattering amplitude $\ampli$ of a dipole of size $\rt$ which, integrated over the transverse plane, gives the total cross  section:
\begin{equation}
\sigmadip(x,\rt)= 
2 \int \ud^2 \bt \ampli(x,\rt,\bt)
\quad ; \quad \int \ud^2\bt e^{i \Deltat \cdot \bt}.
\end{equation}
The predictive power of the dipole framework lies in the fact that from this same degree of freedom one can calculate cross sections for a multitude of processes:
\begin{itemize}
\item 
The total  $\gamma^* p$ or $\gamma^* A$ cross section is obtained by convoluting the the dipole amplitude by the square of the photon-to-$q\bar{q}$ wavefunction:
\begin{equation}
\sigma^{\gamma^*H}_{\textnormal{tot}} = 
\left| \Psi({\gamma^* \to q\bar{q}} )\right|^2 
\otimes \sigmadip
\textnormal{ with }
\otimes \equiv \int_0^1 \ud z \ud^2\rt .
\end{equation}
\item Cross sections for inclusive diffraction, on the other hand, are given by the square of the dipole amplitude, Fourier-transformed from impact parameter to momentum transfer:
\begin{equation}
\frac{\ud \sigma^{\gamma^*H \to X + H}}{\ud t} = 
\frac{1}{4\pi}
\left| \Psi({\gamma^* \to q\bar{q}} )\right|^2 
\otimes
\left|\ampli (\Deltat) \right|^2 .
\end{equation}
\item  Exclusive vector meson cross sections require, in addition to the photon wave function, the light cone wavefunction of the vector meson:
\begin{equation}
\frac{\ud \sigma^{\gamma^*H \to V H}}{\ud t} = 
\frac{1}{4\pi}
\big|
 \Psi({\gamma^* \to q\bar{q}} )
\otimes
\ampli (\Deltat)
\otimes
 \Psi^{*}(q\bar{q} \to V) 
\big|^2 .
\end{equation}
\item  Cross sections for inclusive particle production  in pp, p$A$ and  $AA$ collisions are expressed in terms of the Fourier transform (from the dipole size $r$ to the produced particle momentum) of the same dipole amplitude. 
\end{itemize}



\begin{wrapfigure}{R}{0.4\textwidth}
\centerline{\includegraphics[width=0.4\textwidth]{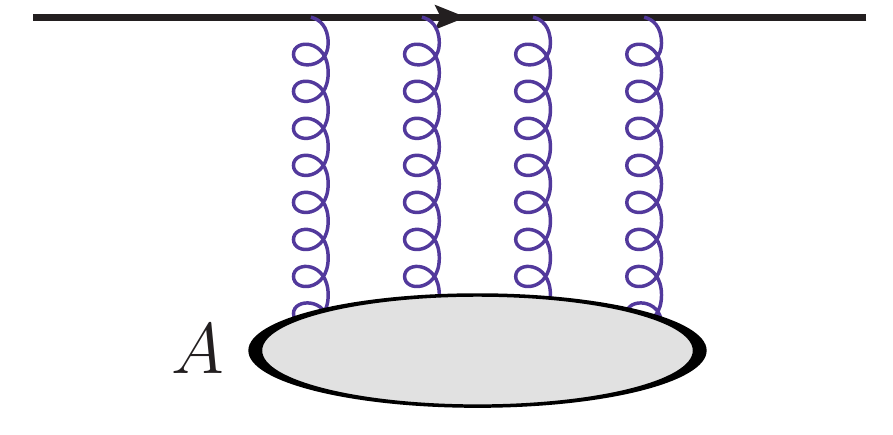}}
\caption{High energy quark interacting with classical gluon field.}
\label{fig:gluonfield}
\end{wrapfigure}
In the CGC~\cite{Gelis:2010nm}
effective theory picture the small-$x$ degrees of freedom of the target are described in as a classical color field. The scattering amplitude for a colored pointlike probe from such a field is given by an eikonal light-like path ordered exponential: the \emph{Wilson line} (see Fig.~\ref{fig:gluonfield})
\begin{equation}
U(\xt) = P \exp\left\{i g \int \ud x^- A^+ (\xt,x^-) \right\}.
\end{equation}
The classical color field is radiated from the larger-$x$ partons of the target. We do not, however, care about the number or nature of these partons, but only about their total color charge (see Ref.~\cite{Dumitru:2021hjm} for recent work reported at this conference).
Thus the amplitude for a color neutral quark-antiquark dipole is given by the trace of a product of a Wilson line at the transverse coordinate of the quark and a conjugate Wilson line at the position of the antiquark, the dipole operator
\begin{equation}
\ampli = 
\frac{1}{\nc}
\mathrm{Tr}
\left\langle 1 -
U^\dag\left(\bt + \frac{\rt}{2}\right)U\left(\bt - \frac{\rt}{2}\right) \right\rangle.
\end{equation}
One of the attractive features of this formulation follows from the fact that the Wilson lines and their products are, by construction, SU($\nc$) matrices that live on a compact manifold. This naturally leads to a bounded amplitude $\ampli \leq 1$ corresponding to gluon saturation built into the formalism.

\begin{wrapfigure}{R}{0.4\textwidth}
\centerline{
\includegraphics[width=5.4cm]{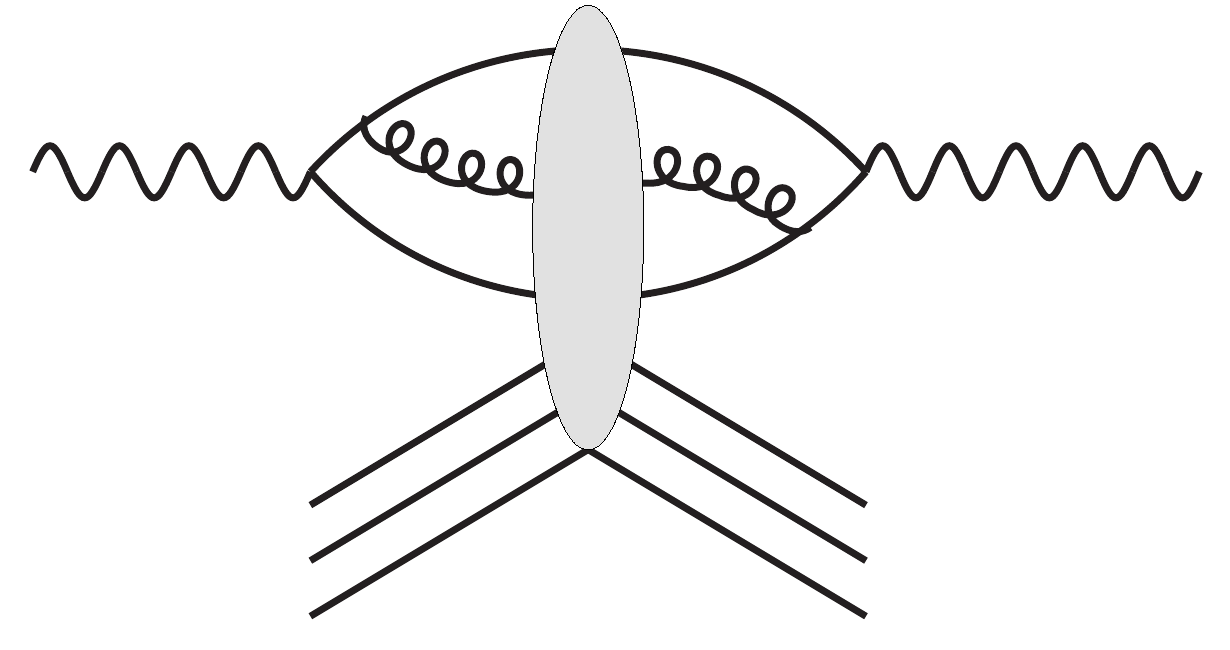}
}
\caption{Adding a gluon to the dipole picture of DIS.}
\label{fig:nlodipole}
\end{wrapfigure}
In addition to providing an intuitive picture of high energy scattering in the saturation regime, the dipole picture is a starting point for a systematical perturbative expansion in weak coupling QCD. Higher order corrections can be calculated by adding gluons to the picture, see Fig.~\ref{fig:nlodipole}. The classical gluon field picture of the target provides a unique way to the calculate the scattering amplitudes of higher Fock states. The phase space region where a gluon is soft generates a contribution that is enhanced by a large logarithm (Leading Log (LL) $\sim \as\ln 1/x$,  Next-to-Leading Log (NLL) $\sim \as^2 \ln 1/x$), which must be absorbed into a renormalization of the target gluon field. This leads to the BK/JIMWLK~\cite{Balitsky:1995ub,Kovchegov:1999yj,
Mueller:2001uk} renormalization group evolution equations that resum these logarithms. The part of phase space of the one-gluon correction that is not enhanced by a logarithm is then a genuine NLO $\sim \as$ correction.

The last years have seen significant advances in pushing this perturbative expansion to the NLL and NLO levels. The NLO versions of the BK~\cite{Balitsky:2008zza} and JIMWLK~\cite{Balitsky:2013fea,Kovner:2013ona} equations have been derived and complemented with collinear resummations needed to stabilize them~\cite{Iancu:2015vea,Ducloue:2019ezk}.
The total DIS cross section for massless quarks has been calculated~ \cite{Balitsky:2010ze,Beuf:2016wdz,Beuf:2017bpd,Hanninen:2017ddy}, and used to obtain a good description of HERA total cross section data~\cite{Beuf:2020dxl}. Diffractive dijet production in DIS has also been addressed~\cite{Boussarie:2014lxa }. The phase space for proper jets at small~$x$ is more limited at the EIC, but these calculations are currently being extended to the diffractive structure functions which have been measured at HERA and certainly will be accessible at the EIC. Similarly one can also calculate exclusive electroproduction of light vector mesons~\cite{Boussarie:2016bkq} using a parton distribution amplitude to describe the meson. Exclusive heavy quarkonium production  is a particularly important observable for QCD, both because of the clean experimental signature~\cite{AbdulKhalek:2021gbh} and because the mass of the heavy quark makes it possible to understand the physics of the bound state in a weak coupling picture. These calculations require virtual photon wavefunctions for massive quarks, which are now gradually starting to become available at NLO~\cite{Beuf:2021qqa}.  Significant advances in pushing the calculations of exclusive quarkonium production to NLO in the dipole picture~\cite{Escobedo:2019bxn,Mantysaari:2021ryb} have been reported at this conference~\cite{Lappi:2021oag}. Simultaneouly there has been a lot of related activity to also move calculations of particle production at forward rapidity in hadron-hardon collisions to NLO accuracy, facing many of the same technical issues as in DIS.

Contrary to the infinite momentum frame partonic picture of the target in collinear factorization, in the dipole picture of DIS one is perturbatively calculating the partonic structure of the virtual photon. 
The most important theory ingredient in these calculations is the photon light cone wave function (LCWF). The LCWF's are defined as the coefficients $\psi^{\gamma\to q\bar{q}},$ $\psi^{\gamma\to q\bar{q}g}$,~\dots of an expansion of the interacting ``dressed'' photon state in terms of ``bare'' Fock states:
\begin{equation}
\left|\gamma^* \right>_D = \left|\gamma \right>_B 
+  \psi^{\gamma^*\to q\bar{q}}  \left|q \bar{q} \right>_B
+ \psi^{\gamma^*\to q\bar{q}g}\left|q \bar{q} g \right>_B+ \dots
\end{equation}
The Hamiltonian perturbation theory calculations in light cone gauge and mixed longitudinal momentum--transverse coordinate space that are needed to calculate the LCWF's could be called the final frontier of QCD perturbation theory. Perhaps as one of the last one-loop perturbative QCD results they are only now becoming fully available, e.g. for the splitting $\gamma^*_{T,L} \to q\bar{q}$ with massless quarks~\cite{Beuf:2016wdz,Beuf:2017bpd,Hanninen:2017ddy}, $\gamma^*_{L}\to q\bar{q}$ and soon $\gamma^*_{T}$  with quark masses~\cite{Beuf:2021qqa}. Similar calculations for the $q\to q g$ LCWF's, needed for hadronic collisions, are also advancing. It is important to realize that there is no ``CGC'' effective theory involved in the LCWF's themselves; they are purely perturbative fundamental quantities describing the elementary vertices. The LCWF's are,  however, crucial \emph{ingredients} of CGC calculations of experimental observables.

As discussed above, calculating exclusive vector meson cross sections requires, in addition to the photon LCWF and the dipole amplitude, a LCWF for the vector meson. Typically in the phenomenological literature one has used simple LCWF parametrizations fit to leptonic decay widths, such as the Boosted Gaussian or LC-Gaus parametrizations~\cite{Kowalski:2006hc}. Improving the precision here requires, however, a more systematical approach. One option is to start from a nonrelativistic potential model wave function and transform it into light cone coordinates (see e.g.~\cite{Krelina:2018hmt}).  One can also use the nonrelativistic limit in a more systematical way by expanding around the limit of large quark mass. This nonrelativistic QCD (NRQCD) approach expresses the bound state properties in terms of universal long distance matrix elements, which can be taken from e.g. decay width data and then used for production cross sections. This is a commonly used approach for inclusive quarkonium production, and is now also becoming available for exclusive DIS~\cite{Escobedo:2019bxn,Lappi:2021oag}.
An approach that is closer to the high energy limit from the start is to directly solve the bound state problem on the light cone, using a phenomenological (e.g. AdS-motivated) confining potential. Starting from Ref.~\cite{Li:2015zda}, there has been a systematical program to develp a set of vector meson LCWF's from such an approach, constrained by a broad set of quarkonium spectroscopic data.

\section{Collinear factorization for exclusive processes}

Exclusive processes can also be addressed from the point of view of collinear factorization. Here one organizes perturbation theory to be able to resum large logarithms of the virtuality $Q^2$ by DGLAP or ERBL renormalization group equations. For exclusive vector meson production in ultraperipheral collisions $Q^2\approx 0$,  but one can argue that the charmonium mass provides a hard enough scale for the formalism to be applicable. Thus the goal is to use photoproduction data to determine the initial conditions for the evolution towards higher $Q^2$.

\begin{wrapfigure}{R}{0.4\textwidth}
\centerline{
\includegraphics[width=5.4cm]{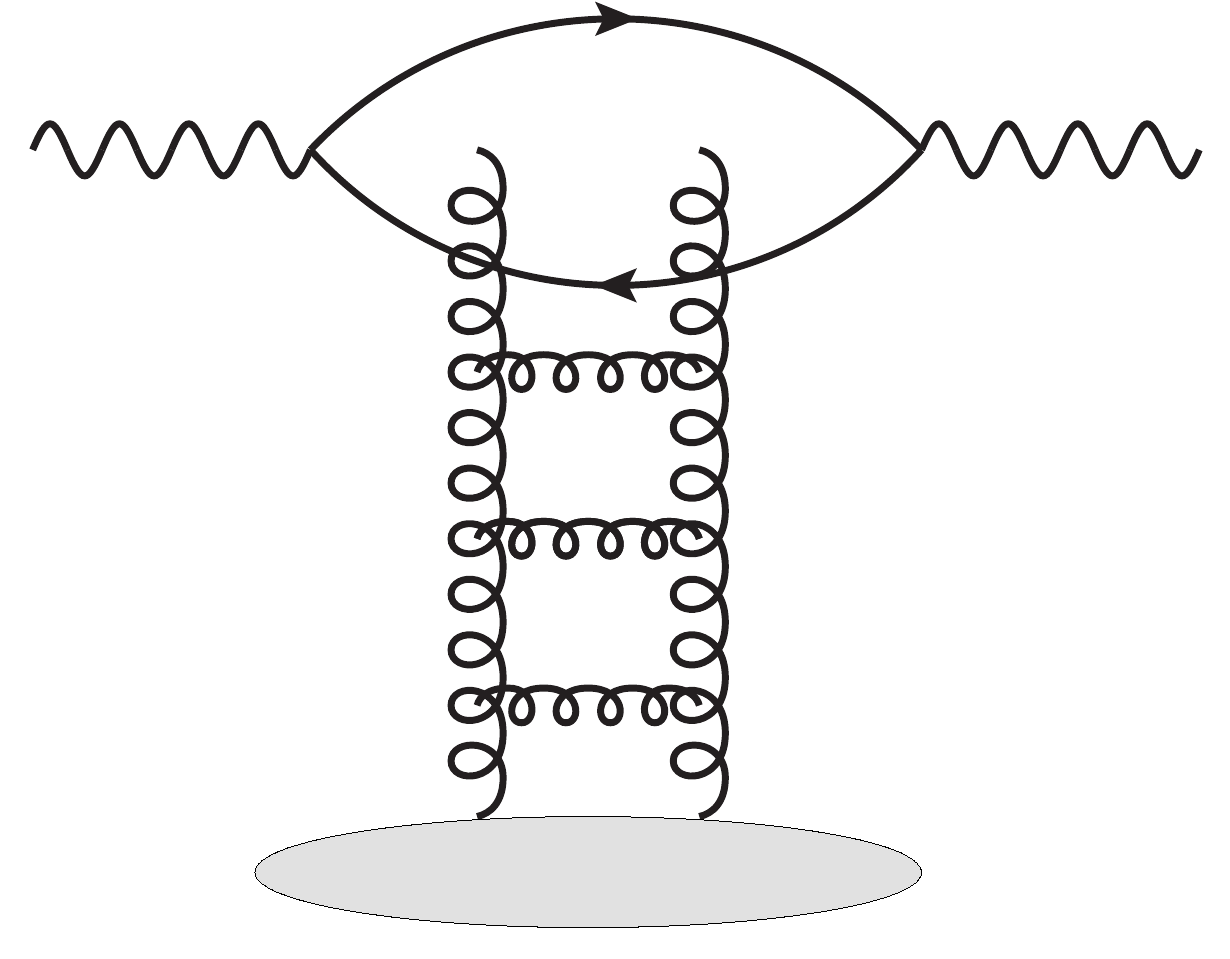}
}
\caption{Dipole amplitude in the two gluon exchange approximation.}
\label{fig:twogch}
\end{wrapfigure}
The collinear and small-$x$ formalisms share a common region of validity. To get an intuitive picture of their relation we can start from the dipole picture and go to the perturbative dilute limit. The dipole amplitude is an elastic amplitude, with a color neutral outgoing state. Thus one exchanges vacuum quantum numbers with the target, i.e. a ``pomeron'' exchange. The simplest such state in a perturbative setup is a 2-gluon exchange in the elastic amplitude, corresponding to the square of a single gluon exchange inelastic amplitude, see Fig.~\ref{fig:twogch}.  Thus, to leading order in  the dilute limit, the dipole amplitudes should be proportional to the gluon distribution. A more careful calculation yields the result~\cite{Frankfurt:1996ri}:
\begin{equation}
\sigmadip(\rt) = 
\frac{\pi^2}{\nc}  \as(\mu^2) xg(x,\mu^2) 
\rt^2 
\end{equation}

\begin{wrapfigure}{R}{0.4\textwidth}
\centerline{
\includegraphics[width=5.4cm]{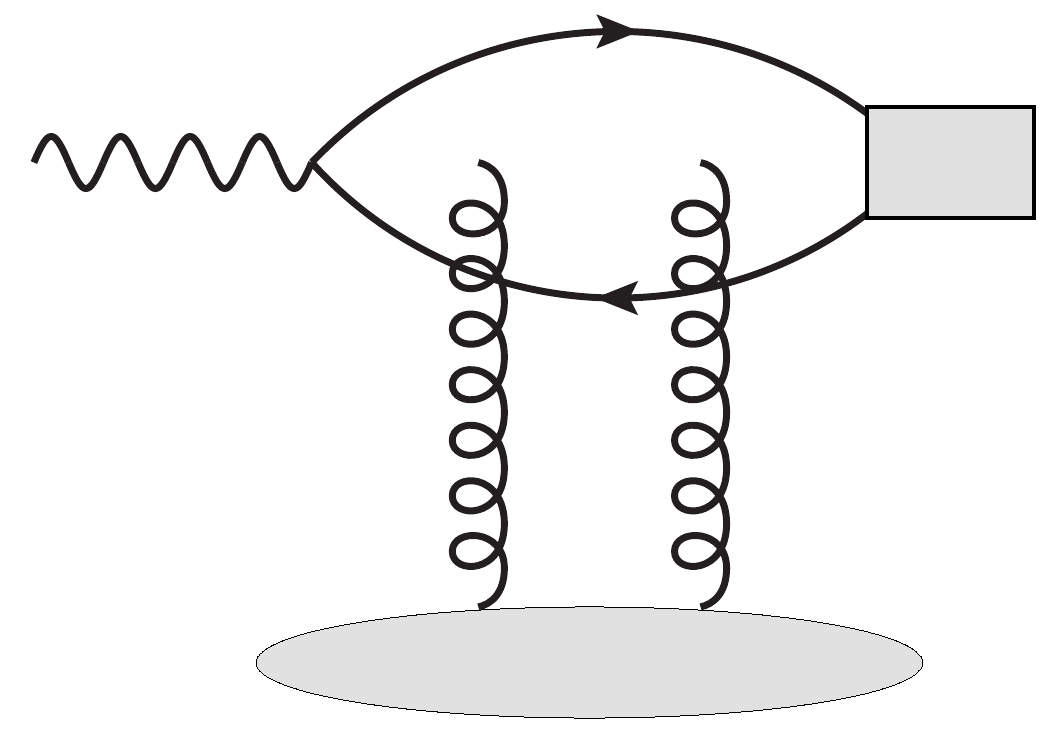}
\begin{tikzpicture}[overlay]
\node[anchor=west] at(-2.9,2.9) {$C_{q/g}$};
\node[anchor=west] at(-2.9,0.33) {$F_{q/g}$};
\node[anchor=west] at(-1.1,2.95) {$\phi_{VM}$};
\node[anchor=west] at(-1.8,1.5) {$(x-\xi)P^+$};
\node[anchor=east] at(-3.2,1.5) {$(x+\xi)P^+$};
\end{tikzpicture}
}
\caption{Exclusive vector meson amplitude in the GPD formulation}
\label{fig:twogchv3}
\end{wrapfigure}
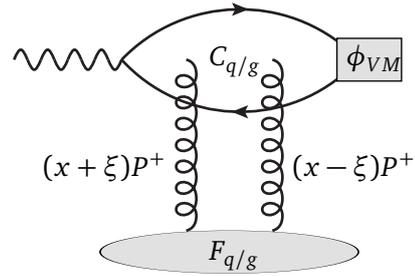
For higher orders in perturbation theory this estimate is clearly not enough. To treat exclusive processes properly in perturbative QCD one needs to introduce the concept of Generalized Parton Distribution (GPD). The GPD is a one-body operator describing the gluon (or quark) content of the target as measured in an off-diagonal transition with a hard external scale, see Fig.~\ref{fig:twogchv3}. In terms of equations one writes the scattering amplitude as a convolution
\begin{multline}
\mathcal{A} \sim  \int_{-1}^{1}\ud x 
\Bigg[
C_g(x,\xi)
F_g(x,\xi)
\\
+ \sum_{q=u,d,s}
C_q(x,\xi)
F_q(x,\xi)
\Bigg] \phi_{VM}.
\end{multline}
Here $C_{g/q}(x,\xi)$ are perturbatively calculable coefficient functions describing the elastic photon-parton scattering. 
The vector meson wave function $\phi_{VM}$ is, in current phenomenological applications, typically taken to be fully nonrelativistic, so that it is just a constant proportional to the leptonic decay width of the meson. The partonic content of the target is described by the GPD's $F_{g/q}(x,\xi)$. The GPD's are collinear distributions. This means that one keeps track of the explicit dependence on the longitudinal momentum fractions
$x+\xi$, $x-\xi$, but the transverse momentum only appears as a renormalization scale. One is thus working in a kinematical regime where the transverse momenta circulating in the diagram are strongly ordered. 
In principle the GPD's are independent quantities that are related to the conventional parton distrbutions only in certain limit. However, based on some relatively general assumptions, in the small-$x$ limit they can be related to the PDF's by the ``Shuvaev transform''~\cite{Shuvaev:1999ce}, which is commonly used in phenomenological applications to extract the gluon PDF from exclusive data.

Taking this approach to NLO accuracy in a phenomenologically reasonable way has been a multi-year research program~\cite{Ivanov:2004vd,Flett:2021ghh}. The NLO corrections are large and strongly dependent on the factorization scale, even to the extent that they can change the sign of the amplitude. However, after a lot of work to stabilize the perturbative expansion, one is now in a position to use exclusive vector meson data to extract the gluon distribution from data at NLO accuracy~\cite{Flett:2020duk}. The exclusive data is not yet included in global PDF fits, but the tools now seem to be settling into place to do so.

\section{Exclusive vector meson measurements}

Measurements of exclusive vector meson cross sections have been a mainstay of the UPC program at the LHC (see e.g. \cite{LHCb:2018rcm,ALICE:2020ugp,ALICE:2021gpt,ALICE:2021jnv}) and at RHIC. Initially the cross sections have been statistics limited, but as the LHC luminosity grows, they are  constantly getting more detailed and differential, helping to differentiate between theoretical models. 

As an example of recent developments is the ALICE measurement~\cite{ALICE:2021tyx} of the coherent $\gamma + A \to J/\Psi + A$ cross section as a function of the momentum transfer  $t$. One thing that is remarkable about this measurement is the access to extremely small values of transverse momenta: the results are reported in a range of $-t\sim 0.0005\dots0.1$GeV$^2$. These values are extracted from the transverse momenta of the $J/\Psi$ decay leptons, since the recoil nucleus cannot be measured. The EIC will have good acess to larger $|t|$ values that are relevant for imaging the proton, but obtaining a comparable resolution at small $|t|$ will be a major challenge. Generally the reported $t$-distribution is steeper than expected from theory. Quantitatively such a development would be expected from saturation, which should make the impact parameter profile flatter. However, the specific saturation model~\cite{Bendova:2020hbb} compared to the experimental data by ALICE is not steep enough to describe it well.

The coherent cross section, where the target does not break up in the interaction, measures the gluon density in the transverse plane averaged over the configurations of the target.
Complementary information about the gluonic structure can be obtained by simultaneously measuring the \emph{incoherent} cross section, where the target breaks up into a relatively low invariant mass color neutral system, separated from the vector meson by a rapidity gap. In the Good-Walker picture the incoherent cross section  measures the fluctuations of the gluon density,
with the value of the momentum transfer $t$ related to the transverse length scale of these fluctuations. A full understanding of the small-$x$  gluonic structure of the target requires measurements and calculations of both coherent and incoherent processes. 

A challenge in separating coherent and incoherent processes with a nuclear target is that the typical momentum transfer in the region where they are both important is quite small, meaning that the recoil nucleus disappears down the beampipe without being detected. The nuclear breakup process itself is also not well understood theoretically. The LHC experiments separate the contributions by fitting the total $t$-distribution to a linear superposition of templates for different processes. At the EIC one will try to veto nuclear breakup events using forward detectors, but doing this accurately enough to get a coherent signal at higher $t$ will be challenging~\cite{AbdulKhalek:2021gbh}.

\section{Other UPC measurements}

Exclusive vector meson production has been the first channel to focus on in ultraperipheral collisions, presumably because it is the easiest one to trigger on. Now, however, the program is extending to more inclusive reactions. A recent intriguing example is the study~\cite{CMS:2020ekd} by the CMS collaboration of the angular dependence of diffractive photon-nucleus dijet events. Here one measures the dijet angular correlation in $\varphi= \varphi_{\mathbf{Q}_T,\mathbf{P}_T}$, the angle 
 between the sum $\mathbf{Q}_T = \pt_1+\pt_2$ and difference $\mathbf{P}_T = \frac{1}{2}\left(\pt_1-\pt_2\right)$ of dijet momenta, where $\pt_1,\pt_2$ are the transverse momenta of the jets. This observable provides access to the the physics of gluon angular momentum, through the gluon transverse momentum distribution (TMD). Such processes will be a major part of the physics program for the EIC,  however, in a much more limited kinematical phase space.

Another slightly more exotic avenue for the audience of this conference is the study of what could be termed  ``non-QCD final states''. A photon-photon interaction in an ultraperipheral collision provides the highest energy photon-photon collisions currently available. One application are the first measurements of rare but conventional QED processes. For example the STAR measurement~\cite{STAR:2019kxd} of angular modulations in $\gamma+\gamma \to \ell^+ \ell^-$ confirms QED predictions for the polarization structure and $\ktt$-dependence of the photon flux, also serving as a normalization check for QCD studies. The direct measurement of  Light-by-light scattering by ATLAS~\cite{ATLAS:2020hii} provides not only a direct test of a highly suppressed one-loop effect of QED, but a way to search for new axion-like particles.

As a first example of an interesting QCD analysis in  \emph{inclusive} photon-mediated processes, the ATLAS collaboration has reported preliminary results on inclusive dijet production in photon-nucleus collisions~ \cite{ATLAS:2017kwa}. This analysis is unfortunately still not finalized and published. If it reaches a good enough level of accuracy, it would serve as a theoretically controlled and sensitive probe of small-$x$ gluon distributions~ \cite{Guzey:2018dlm}.

As discussed in the earlier sections, at high energy the photon state  also has an important hadronic component, which becomes important in high energy scattering. In the absence of a hard scale, i.e. for photoproduction and without heavy quarks, this means that  effectively the photon starts to act like a $\rho$-meson. One can then see it as a lighter version of a proton, and use heavy-ion-like analysis techniques that have found new areas of application in smaller collision systems, proton-nucleus and high multiplicity proton-proton collisions. The archetypical such analysis is to look for ``flow'' structures in azimuthal multiparticle correlations that, at least in a heavy ion event, are caused by hydrodynamical final state interactions. The ATLAS  collaboration has performed precisely such an analysis~\cite{ATLAS:2021jhn}, dividing photon-nucleus events into multiplicity classes. In the high multiplicity event sample one identifies a $\cos 2 \varphi$ structure that could be interpreted as the signal of final state interactions, a tiny droplet of quark gluon plasma created in a photon-induced collision.

\section{Conclusions}

The major part of this contribution has discussed  recent advances in small-$x$ theory in the gluon saturation regime. We have argued that there has been a systematical push to move to NLO accuracy in perturbative weak coupling calculations of DIS and other dilute-dense processes. To consistently include the effects of gluon saturation, these calculations are preferrably performed in a mixed longitudinal momentum-transverse coordinate space. The interaction with the target is described by eikonal scattering amplitudes, Wilson lines.
Simultaneously, also in the complementary collinear factorization picture the description of exclusive reactions and their connections to inclusive ones has been moving to NLO accuracy, including in  phenomenological applications. 

While a lot of this theory development is motivated by the EIC physics program, we do not need to wait for the EIC for exciting new measurements. The LHC and RHIC have a growing and versatile program of photon-mediated ultraperipheral collisions that addresses the same physics topics. Ultraperipheral collisions reach higher collision energies than the EIC will, but only involve quasi-real photons. Thus they require a heavy quark or a jet to stay in the weak coupling regime. The EIC will add to this a leverarm in $Q^2$ that turn new kinds of observables into perturbative probes of gluons in ordinary baryonic matter. Thus there is a strong complementarity between the ultraperipheral and EIC programs. 

\section*{Acknowledgements}

 This work has been supported by the European Research Council under grant no.~ERC-2015-CoG-681707,  by the EU Horizon 2020 research and innovation programme, STRONG-2020 project (grant agreement No 824093) and by the Academy  of  Finland,  project 321840.   The content of this article does not reflect the official opinion of the European Union and responsibility for the information and views expressed therein lies entirely with the authors.



\bibliography{dis21procs.bib}

\nolinenumbers

\end{document}